# Infrared thermochromic antenna composite for self-adaptive thermoregulation

Francisco V. Ramirez-Cuevas[1,2†], Kargal L. Gurunatha[1,3†], Lingxi Li[1], Usama Zulfiqar[1], Sanjayan Sathasivam[4,5], Manish K. Tiwari[6], Ivan P. Parkin[5], Ioannis Papakonstantinou[1*]

[1]Photonic Innovations Lab, Department of Electronic & Electrical Engineering, University College London; London WC1E 7JE, United Kingdom

[2]Center for Energy Transición (CENTRA), Facultad de Ingeniería y Ciencias, Universidad Adolfo Ibáñez; Santiago 7941169, Chile

[3]Centre for Nano and Material Science (CNMS), JAIN University; Ramanagara Bangalore 562112, India

[4]School of Engineering, London South Bank University; London SE1 0AA, UK

[5]Materials Chemistry Centre, Department of Chemistry, University College London; London WC1H 0AJ, UK

[6]Nanoengineered Systems Laboratory, Department of Mechanical Engineering, University College London; London WC1E 7JE, UK

*Corresponding author. Email: i.papakonstantinou@ucl.ac.uk

† These authors contributed equally to this work

**Self-adaptive thermoregulation, the mechanism living organisms use to balance their temperature, holds great promise for decarbonizing cooling and heating processes. The functionality can be effectively emulated by engineering the thermal emissivity of materials to adapt to background temperature variations. Yet, solutions that marry large emissivity switching ($\Delta\epsilon$) with scalability, cost-effectiveness and design freedom are still lacking. Here, we fill this gap by introducing infrared dipole antennas made of tunable thermochromic materials. We demonstrate that non-spherical antennas (rods, stars and flakes) made of vanadium-dioxide can exhibit a massive (~200-fold) increase in their absorption cross-section as temperature rises. Embedding these antennas in polymer films, or simply spraying them directly, creates free-form thermoregulation composites, featuring an outstanding $\Delta\epsilon \sim 0.6$ in spectral ranges that can be tuned at will. Our research paves the way for versatile self-adaptive heat management solutions (coatings, fibers, membranes and films) that could find application in radiative-cooling, heat-sensing, thermal-camouflage, and other.**

Self-adaptive thermoregulation is the mechanism living organisms use to maintain their body temperature stable against thermal fluctuations of their surroundings. Researchers have long sought imitating this mechanism to dynamically balance heat influx and outflux in their systems and dampen detrimental temperature swings [1–6]. An efficient way to regulate the temperature of a surface is by controlling the amount of heat it is allowed to radiate. In that respect, both passively and actively controlled radiators have been developed, whose emissive power is tuned by some external stimulus (temperature, electrical current, mechanical actuation or other) [3,7–12].

Passive systems are of particular interest as they do not consume energy during operation. Much of the work in this area has focused on thermochromic materials, i.e. materials undergoing a temperature driven phase-change around a critical temperature ($T_c$)[3,6]. Several materials have been used for this purpose including metal-oxides (e.g. $VO_2$) [7,8,13], chalcogenides [e.g. $Ge_2Sb_2Te_5$ (GST) and $In_3SbTe_2$ (IST)] [17,18], perovskites [e.g. $SmNiO_3$ and $(La,Sr)MnO_3$ (LSMO)] [11,19–21], liquid crystals and other [3,6]. Temperature-driven heat emission is corollary to thermochromism by way of Kirchoff's law of thermal radiation. This is due to the absorbance of thermochromic materials constantly changing throughout the phase-change,. As a result, a continuously varying emissivity profile manifests, bounded between its asymptotic values at low and high temperatures (also known as cold and hot states). The design variable to maximize is naturally the hot-to-cold emissivity contrast $\Delta\epsilon = \epsilon_h - \epsilon_c$ ($\epsilon_{h/c}$, emissivity in the hot/cold states). To achieve passive thermoregulation, it is essential that $\Delta\epsilon > 0$ (positive differential), [7,8,13,14]. For when the thermochromic medium deviates from its transition temperature, a radiative feedback develops forcing it back towards $T_c$. The spectral range for maximizing $\Delta\epsilon$ is usually set to either the Long Wavelength Infrared (LWIR) atmospheric transparency window ($8 - 13$ μm), or the Mid-Infrared (MIR) window ($3 - 5$ μm). The former aligns with the peak of the blackbody distribution at temperatures between $25 - 100$ °C, the latter is commonly used in thermal imaging applications [15,16].

The most common approach in the literature to maximize $\Delta\epsilon$ is based on Fabry-Perot cavities., Several variants of this approach have appeared utilizing multilayered [7,22–24] or metasurface [8,14,25–28] or other resonant structures [13, 29–30]. Whilst some of these concepts have demonstrated adequate emissivity switching, scaling them up is a major challenge. A number of them, for example, require specialized equipment and cleanroom environments that increase the final cost [7,8,14,27,31]. Furthermore, their fabrication typically requires hard substrates (silicon, quartz, sapphire or other) to withstand the elevated temperatures during deposition, or post annealing steps and to ensure uniformity across the whole area [7,14,27,31]. Permanently attached substrates though, impose constraints making them incompatible with anything but planar surfaces. A recent study utilized thermochromic-shelled particles of spherical geometry to overcome some of these issues [13]. Unfortunately, only modest emissivity modulation ($\Delta\epsilon \sim 0.4$) was demonstrated. This is because, as we show later, core-shell spherical geometries are not optimum to maximize the composite's emissivity. They are also not the best design to economize on material usage.

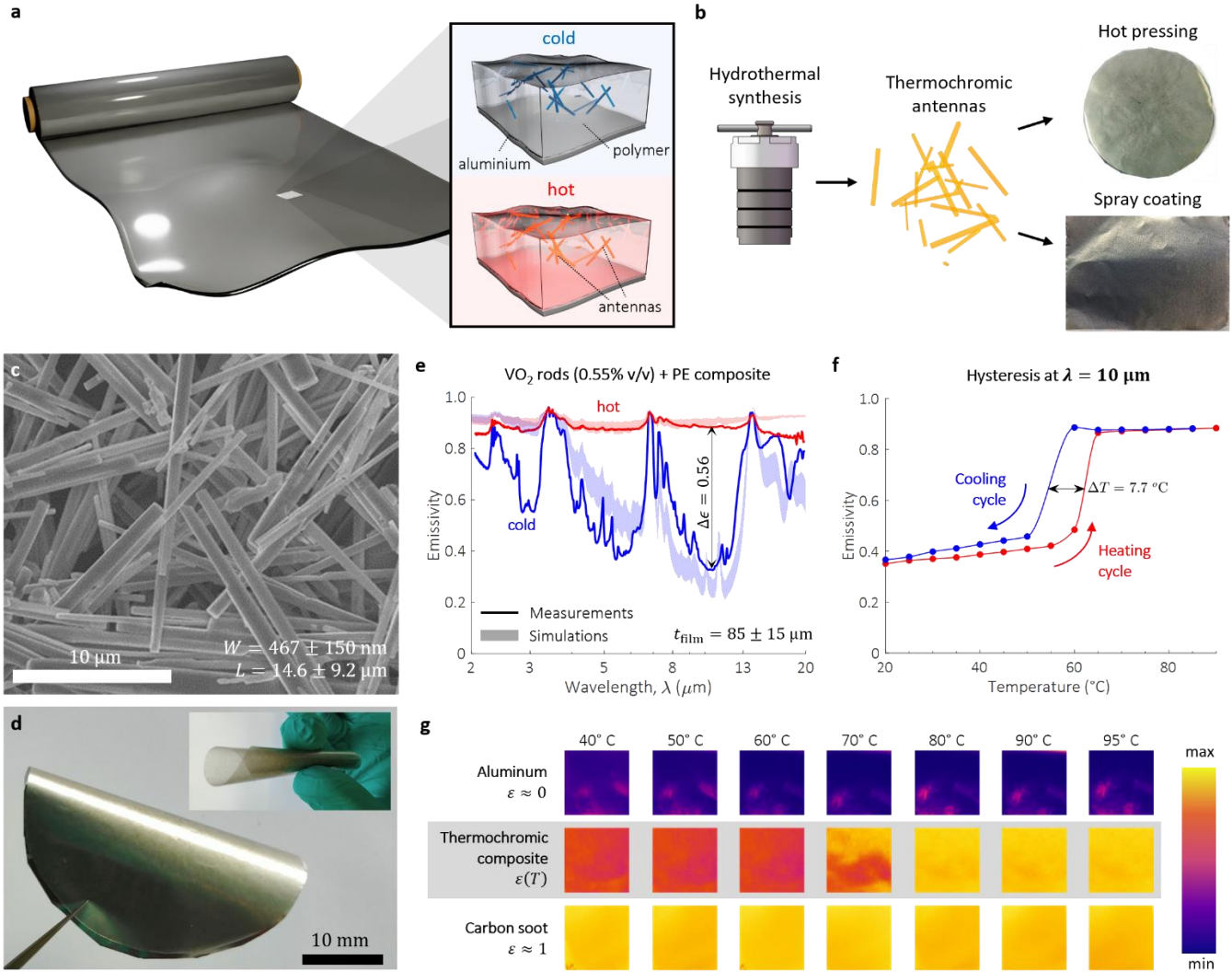

**Fig. 1. Thermochromic antennas for passive thermoregulation composites. a.** Concept of self-adaptive thermoregulation composite film consisting of infrared thermochromic antennas embedded in a polymer matrix. **b.** Schematic of the fabrication protocol for free- composites. VO₂ antennas were first synthesized by hydrothermal synthesis in a high-pressure autoclave. Polymer films and coatings were then made by hot pressing and spray coating respectively [33]. **c.** Scanning electron microscopy (SEM) image of VO₂ antennas, showing the width ($W$) and length ($L$) size distribution. **d.** Photograph of composite thermochromic film made by mixing VO₂ antennas with polyethylene (0.55% v/v) and hot-pressed against an aluminum foil. The polymer film composite has a thickness of only $85 \pm 15$ μm and is flexible (inset). **e.** Experimental characterization of the hot/cold emissivity of the composite film (solid line). The red and blue filled area correspond to the results from simulations [33]. **f.** Hysteresis loop at $\lambda = 10$ μm during heating and cooling cycles. **g.** Thermal camera images of carbon black ($\epsilon \approx 1$), aluminum ($\epsilon \approx 0$) and thermochromic composite heated from 40°C to 95°C. The switch at ~70°C of the composite from a low to a high emissivity state is evident.

Departing from the state-of-the-art, we propose an alternative route to create efficient self-adaptive heat radiators by finetuning the absorption cross-section in infrared thermochromic antennas (Fig. 1a-b). In particular, we establish that: i) Dipole antennas with high surface area-to-volume ratio (SA:V) is key to achieve large absorption cross-section ratios, of which rod geometries (Fig. 1c) are the best performing. ii) When dispersed in a host material, the collective action of the antennas macroscopically translates into a composite medium with large emissivity switching (Fig. 1e). iii) Only a very small amount of antennas is sufficient to attain large $\Delta\epsilon$. The medium hosting them can be ultrathin (<100 μm) and flexible (Fig. 1d) with obvious benefits in terms of cost and design freedom. iv) Antenna resonances are very sensitive to their geometrical dimensions, a feature that can be exploited for wavelength selective radiators. As proof-of-concept material, we use vanadium-dioxide (VO₂), an archetype, strongly correlated thermochromic metal-oxide with a temperature driven Insulator-to-Metal Transition (IMT) [32]. Rod-shaped VO₂ antennas were made by hydrothermal synthesis (Fig. 1c), a process allowing accurate control of their length and aspect ratio by carefully tuning the reaction stoichiometry, temperature, pH, and post annealing of the reaction products (Materials and Methods) [33]. The antennas showed uniform morphology, with a width ($W$) × length ($L$) size distribution of $(0.47 \pm 0.01$ μm$) \times (14.6 \pm 9.2$ μm$)$. The powder X-ray diffraction (PXRD) patterns and XPS analysis of the final product confirm the VO$_2$ monoclinic (M) phase formation without any impurities, while differential scanning calorimetry (DSC) measurements revealed $T_c = 70.2$ °C for rods and similar trend with stars (Fig. S7). Flexible composite films were subsequently fabricated by hot-pressing a dry mix of high-density polyethylene (HDPE) powder and VO$_2$ antennas over an aluminum film, the latter acting as a back reflector to prevent radiative thermal exchange with the underlying body (Fig. 1d). The composite film exhibited a dramatic emissivity switching at wavelength, $\lambda = 10.5$ μm (Fig. 1e), from $\epsilon_c = 0.327$ to $\epsilon_h = 0.883$ ($\Delta\epsilon = 0.56$), and a narrow hysteresis width, $\Delta T_{\text{hyst}} = 7.7°$ C (Fig. 1f). With a thickness of < 100 μm, the film was also fully flexible (Fig. 1d). First-principle radiative transfer calculations showed very good agreement with experimental results, (Fig. 1f and Materials and Methods). We fabricated and tested other samples to ensure repeatability, rendering similar results in terms of $\Delta\epsilon$ and $\Delta T_{\text{hyst}}$ (Fig. S2). To further assess the thermo-responsive mechanism, we imaged our sample with a LWIR camera [33] and compared against two references; one with constant $\epsilon \approx 0\%$ (aluminium) and another with $\epsilon \approx 100\%$ (Carbon black). The transition of our composite from a low to a high emissivity state at $T \sim 70°$ C is clear (Fig. 1g). Thermochromic antennas are a very versatile starting material from which several end-products may be derived. To prove this point, we developed a second proof-of-concept, whereby a solution of VO$_2$ antennas dispersed in acetone was created that could easily be sprayed directly on any surface (Fig. 1b and S4).

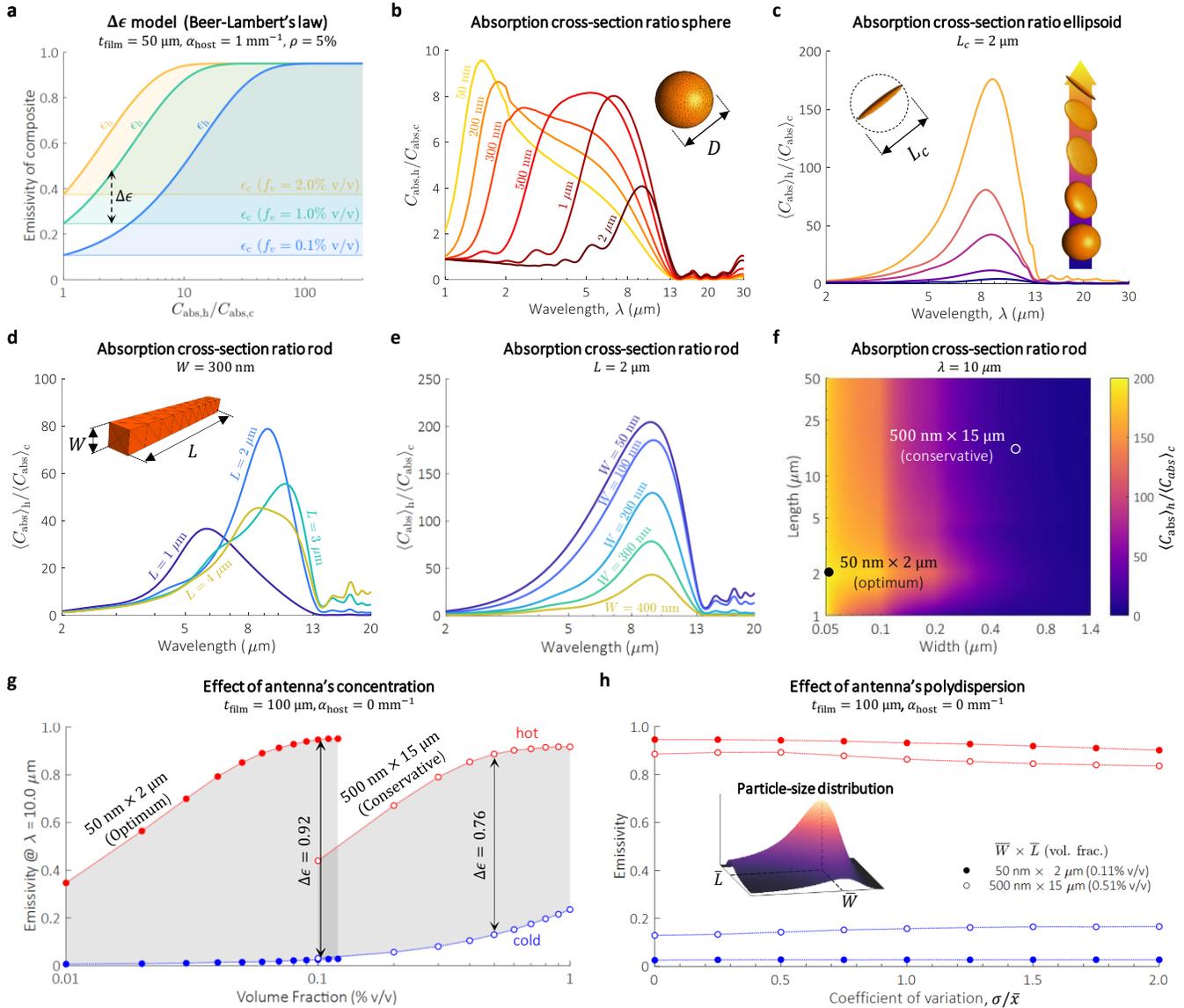

**Figure 2. Computational analysis of thermochromic VO₂ antennas and passive thermoregulation composite. a.** Emissivity contrast of a composite of thickness $t_{film} = 50$ μm as a function of the antenna absorption cross-section ratio (section S2.4). The results are obtained for $C_{abs,c}/V_p = 0.2$ μm$^{-1}$, an absorption coefficient of the host matrix, $\alpha_{host} = 1$ mm$^{-1}$, and reflectance of the composite, $\rho = 5\%$, which represent typical values observed in our experiments. **b.** Absorption cross-section ratio of VO₂ spheres as a function of their diameter, $D$. **c.** Absorption cross-section ratio of VO₂ ellipsoids as function of their surface-area-to-volume (SA:V) ratio. Two of the three principal axes of the ellipsoid are reduced to a length of 100 nm, while the other axis is kept at constant length, $L_c = 2$ μm. The absorption cross-section ratio is based on orientation-averaged values $\langle C_{abs} \rangle$. **d.** and **e.** Spectral absorption cross-section ratio as a function of the rod length ($L$) and width ($W$), respectively. The corresponding absorption cross-section of the hot and cold phases is shown in Fig. S11. **f.** VO₂ rod absorption cross-section ratio at $\lambda = 10$ μm. The results were computed with the $\langle C_{abs} \rangle$ at hot and cold phases shown in Fig. S11. **g.** Hot/cold phase emissivity of a VO₂ rod composite as a function of the volume fraction, considering optimum (filled circles) and conservative (open circles) rod dimensions, the later based on the dimensions of our synthesized antennas. The phase transition region is highlighted by the grey area. **h.** Similar to (g), but with $\langle C_{abs} \rangle$ weighted by a normal particle-size distribution with a variable coefficient of variation $\sigma/\bar{x}$, where $\sigma$ is the standard deviation and $\bar{x}$ is the mean value [33]. In all simulations, the refractive index ($n + i\kappa$) of the host material is 1.5. For VO₂, $n + i\kappa$ is reported elsewhere [32].

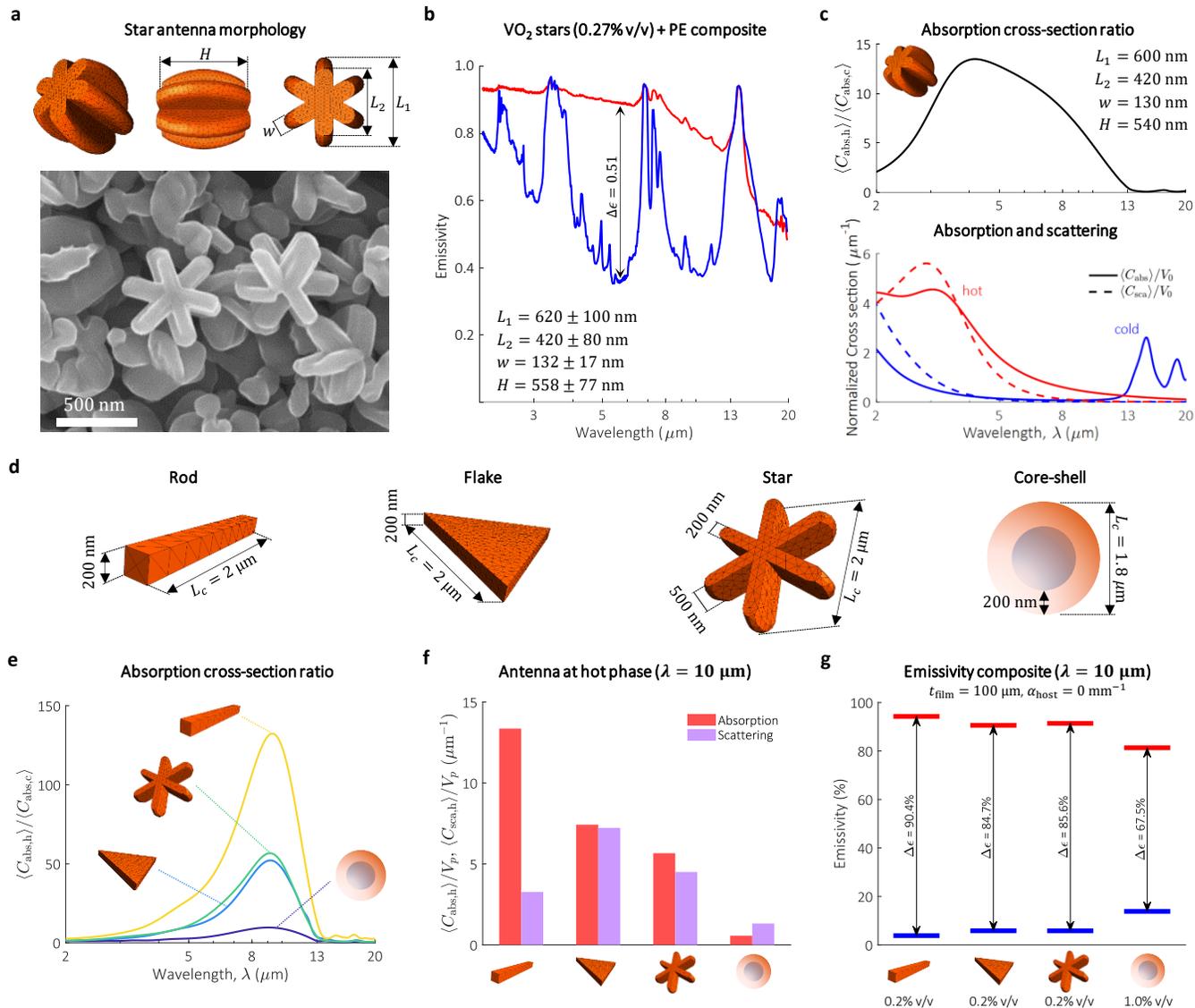

**Figure 3. Role of morphology in the performance of thermochromic antennas. a.** 3D model used in computational calculations (top) and SEM (bottom) of VO$_2$ nanostars **b**. Experimentally measured emissivity of a thermochromic composite based on VO$_2$ stars at hot and cold states. **c.** Simulation of absorption cross-section contrast, and normalized scattering and absorption cross-sections of optimum VO$_2$ nanostars (dimensions shown at the top right corner). **d.** 3 different dipole antenna geometries with large SA:V, and core-shell particle, with their optimized characteristic length ($L_c$). **e.** Absorption cross-section contrast for the 4 different geometries presented in d, **f.** Normalized scattering and absorption cross-sections of the 4 geometries in the hot state at $\lambda = 10$ μm. **g.** Hot/cold emissivity of resulting composites at $\lambda = 10$ μm. In all simulations the refractive index of the host was $N_h = 1.5$. For the core-shell particle, the refractive index of the core was $N_c = 1.5$. For the results shown in (g), $t_{\text{film}} = 100$ μm. The optimum antenna volume fraction $f_v$ is also shown below each label in the x-axis (details in section S2.6)

**Optimization pathways for thermochromic antenna composites**

To maximize $\Delta\epsilon$, the critical design parameter is the antenna's absorption cross-section ratio, $C_{\text{abs,h}}/C_{\text{abs,c}}$, where $C_{\text{abs,h}}(C_{\text{abs,c}})$ is the absorption cross section in the hot(cold) phase. This is illustrated in figure 2a, using a simplified model based on Beer-Lambert's law (Supplementary Material). This model correlates the $C_{\text{abs,h}}/C_{\text{abs,c}}$ ratio of individual antennas with the total emissivity $\epsilon_{\text{h/c}}$ of the composite. As evidenced in Fig. 2a, the larger the emissivity in the cold state $\epsilon_c$, the larger the absorption cross-section ratio required to maximize $\Delta\epsilon$. This helps explain why spherical particles can only ever achieve meagre $\Delta\epsilon$, even though they are some of the most popular geometry in the VO$_2$ literature [7,34,35]. As seen in Fig. 2b, in the range $\lambda \in 8 - 13$ µm the best $C_{\text{abs,h}}/C_{\text{abs,c}}$ that can be achieved is $\sim$4 for optimized VO$_2$ spheres of 2 µm in diameter. For such modest absorption cross-section ratios though, $\Delta\epsilon$ is bounded to $< 0.35$. This situation changes starkly when SA:V increases (Fig. 2c). In the figure, SA:V is increased by deforming the sphere down to a disk and then to a rod, leading to a monotonic enhancement of the absorption cross section ratio. It is important to note that for non-spherical particles, it is the orientation and polarization-averaged absorption cross-sections, $\langle C_{\text{abs,c/h}} \rangle$, that needs to be considered, which physically represent the average response of the randomly-oriented antennas in the composite [36]. In the hot phase, an increase in SA:V suppresses the non-radiative modes, resulting in an increase of radiative dissipation and hence an enhancement of $\langle C_{\text{abs,h}} \rangle$ [37,38]. This is further confirmed by mode decomposition analysis (Fig. S10), which reveals that the absorption cross section is dominated by the contribution of dipole modes as SA:V increases. On the other hand, the absorption in the cold phase is primarily governed by polarization currents rather than free electrons [39]. At wavelengths $\lambda \in 5 - 13$ µm, cold-phase VO$_2$ features only weak polarization currents [32]. and $\langle C_{\text{abs,c}} \rangle$ is proportional to the volume of the structure. The structures analyzed in Fig. 2c for example, feature $\langle C_{\text{abs,c}} \rangle/V_p \approx 0.2$ µm$^{-1}$ (Fig. S9). As a result, $\langle C_{\text{abs,c}} \rangle$ now reduces as SA:V increases. The net result is a sharp increase to the absorption cross-section ratio, which for a rod may exceed 180, or almost 50 times larger than the spherical case.

The magnitude and peak of the absorption cross-section ratio can be finetuned by adjusting its width, $W$ (Fig. 2d), and length, $L$ (Fig. 2e), respectively. For example, setting a target at $\lambda = 10$ µm, which is in the middle of the LWIR window, we identify that the largest absorption cross-section ratio is reached for antennas with $L \approx 2$ µm and $W \approx 50$ nm (Fig. 2f). By reducing the length of the antenna to $L \approx 1$ µm, the peak of the absorption cross-section ratio is shifted to $\lambda \sim 5$ µm, on the other hand. As revealed by first principle radiative transfer simulations (Fig. 2g), the emissivity contrast of a composite based on a transparent host ($\alpha_{\text{host}} = 0$ mm$^{-1}$) and these optimum antennas can reach an impressive $\Delta\epsilon = 0.92$ at an antenna volume fraction of just $f_v = 0.11\%$. In a more conservative scenario based on the dimensions of our synthesized antennas, this limit decreases to $\Delta\epsilon = 0.76$ at $f_v = 0.51\%$ v/v. In both cases, however, only a very small concentration of antennas is required to create an efficient self-adaptive heat radiation system. Interestingly, the predicted $\Delta\epsilon$ shows little sensitivity to polydispersity (Fig. 1h). Considering a normal particle-size distribution with a coefficient of variation of 2, $\Delta\epsilon$ is reduced only by $\sim 5\%$ and $11\%$ for the optimal and conservative scenario, respectively. As a reference, the coefficient of variation of our synthesized antennas is $< 0.63$. It is important to note here that our design rules are universal and applicable on any thermochromic material exhibiting an IMT. For example, maximum $\Delta\epsilon = 0.84$ and 0.92 are predicted for GST and IST rod composites, respectively (Fig. S12).

**Effects of the morphology of thermochromic antennas**

Attaining large absorption cross-section ratio is by no means bound to antennas of rod geometry only, adding additional degrees of design freedom. An effective way to tune the geometry of an antenna is by changing the concentration of surfactant (CTAB in our case), which preferentially adsorbs in specific crystallographic facets during the hydrothermal synthesis [33,40,41]. To demonstrate the versatility of this approach, submicron VO$_2$ nanostars were synthesized, whose emissivity switching was this time optimized for the MIR window (Fig. 3a). When embedded into a PE matrix film, the VO$_2$ nanostar composite exhibited a $\Delta\epsilon = 0.52$ at $\lambda = 5.4$ µm (Fig. 3b). This blue shifting to the maximum of the $\Delta\epsilon$ is attributed to the smaller dimension of the nanostars and agrees with our modelling results (Fig. 3c).

We further explore the role of geometry, by modelling and comparing 4 different structures with different SA:V: stars, flakes, rods, and a spherical core-shell particle composed of a LWIR transparent core and a thin VO$_2$ shell (Fig. 3d-g). Core-shell particles, have been proposed before in the context of emissivity switching and it is interesting to assess how they compare against our dipole antennas [13]. To make the comparison fair, all structures were designed with a moderate width (or shell-thickness) of 200 nm and a characteristic length ($L_c$) optimized to exhibit an absorption cross-section maximum at $\lambda = 10$ μm. The VO$_2$ shell thickness agrees with the dimensions of synthesized CaF$_2$@VO$_2$ core@shell particles [13]. The heigh of the star ($H = 500$ nm) is based on experimentally reported stars of similar length scales ($L_c = 2$ μm) [42]. As shown in Fig. 3e, rods show the largest absorption cross-section ratio (132.4), followed by stars (56.7), flakes (52.1) and core-shell spheres (9.7). So far, the impact of scattering in the optical properties of the composite films has not been discussed. However, since the critical dimensions of the antennas are comparable with the wavelength of thermal emission, scattering may play an important role. The effect of scattering is accounted for by comparing the (orientation-averaged) absorption and scattering cross sections, $\langle C_{\text{abs}} \rangle/V_p$ and $\langle C_{\text{sca}} \rangle/V_p$, normalized to the volume of the antenna (Fig. 3f). The normalized cross-sections calculated in Fig. 3f were then used as input to solve the full radiative transfer equation and calculate the emissivity of the composite films (Fig. S13). As seen in Fig. 3g, scattering might either enhance or reduce $\Delta\epsilon$. If $\langle C_{\text{sca}} \rangle/V_p$ is smaller than $\langle C_{\text{abs}} \rangle/V_p$, scattering plays a favorable role to enhance $\epsilon_h$ as it increases the optical path in the film.[43] This is evidenced in the results of composites with stars and flakes, whose $\epsilon_h$ is similar to those of rod composites, despite having lower $\langle C_{\text{abs}} \rangle/V_p$. Using Beer-Lamberts law (Fig. S13), we confirmed that the enhancement of $\epsilon_h$ disappears when the effect of scattering is not present. If $\langle C_{\text{sca}} \rangle/V_p$ is larger than $\langle C_{\text{abs}} \rangle/V_p$, however, the reflectance of the composite increases, lowering $\epsilon_h$; a feature observed in the results of core-shell spheres. An important point to note is that the magnitude of $\langle C_{\text{abs}} \rangle_h/V_p$ correlates directly to the antenna concentration $f_v$ required to maximize $\Delta\epsilon$. In this regard, rods, flakes and stars composites require at least 5 times less antennas than the core-shell composite.

**Summary**

In summary, we introduced a new family of non-spherical, dipole thermochromic antennas featuring high absorption cross-section contrast that can be leveraged to construct effective self-adaptive radiators. Antennas can be synthesized in bulk using cheap precursors and then mixed with polymers or other materials lending our technology compatible with numerous scalable, industrial manufacturing processes such as film/foil extrusion, fiber extrusion, compression molding, injection molding, dip coating, spray coating, doctor blading, electrospinning and other. In essence, with the right combination of host material and fabrication method, passive thermoregulation products of arbitrary shape and mechanical properties (flexible, rigid, stretchable, twistable, etc.) may be created that can be applied on, virtually, any surface. Our method is also very cost effective. To put this in perspective, a typical hydrothermal reaction in our lab scale autoclaves was sufficient to produce >200 mg of material, when <2.15 mg of rods is required per square meter of composite (Fig. S1). By exploring the effects of morphology, we demonstrate that antenna resonances are very sensitive to their geometrical dimensions, a feature that can be exploited for wavelength sensitive radiators that switch only in certain spectral bands. For example, monodispersed VO$_2$ stars and rods can be used to cover three different spectral responses, respectively, i) a radiator switching in the 3-5 μm MIR window, ii) a radiator switching in the LWIR 8-13 μm window, and iii) a broadband radiator responding across both windows. Interestingly, targeting structures with large SA:V can also positively affect the nature of the thermochromic transition. As observed in our experiments, rods and stars tend to grow along well-defined crystallographic planes with high crystallinity [41], resulting in antennas with sharp IMT transition and narrow hysteresis (Fig. 1f), a desirable feature for some applications such as tunable radiative cooling. Alternatively, $T_c$ of VO$_2$ can be tuned by W-doping [44], or by combined effect of strain and oxygen vacancies [41] for near room-temperature switching. While rods, stars and flakes designs were investigated in this article, there is no indication these are exclusive. Our design rules are universal extending to other morphologies with large SA:V, as well as other thermochromic materials exhibiting IMT transitions. Furthermore, while only designs with $\Delta\epsilon > 0$ have been discussed, if the antenna concentration exceeds the percolation threshold, this situation may be reversed leading to composites with $\Delta\epsilon < 0$[45] (negative differential). In this case, radiative heat losses may stay constant, or even be reversed as the temperature increases [15,46,47]. The apparent temperature of a surface, recorded by an infrared camera, may then be decoupled from its real temperature, creating a thermal camouflage effect [15,16,48].

# Supplementary Materials for
# Infrared thermochromic antenna composite for self-adaptive thermoregulation

Francisco V. Ramirez-Cuevas†, Kargal L. Gurunatha†, Lingxi Li, Usama Zulfiqar, Sanjayan Sathasivam, Manish Tiwari, Ivan P. Parkin, Ioannis Papakonstantinou*

Table of Contents

List of Figures

**Figure S1. a1.** and **b1.** SEM images of as-synthesized VO$_2$ rods and stars used in the results of the main text, respectively. **a2. and a3.** Measured emission spectra of VO$_2$ rods + PE composites based on the sample shown in a1, and volume fractions of 0.14% and 0.27% v/v, respectively. **b2 and b3.** Measured emission spectra of VO$_2$ stars + PE composites based on the sample shown in b1, and volume fractions of 0.14% and 0.55% v/v, respectively. ...............................................................10

**Figure S2. a.** SEM images of another batch of VO$_2$ rods, synthesized according to the guidelines outlined in Materials and Methods (section S1.1). **b. and c.** Measured emission spectra of VO$_2$ rods + PE composites based on the sample shown in a, and volume fractions of 0.27% and 0.55% v/v, respectively. .......................................................................................................................10

# S1. Materials and methods

## S1.1. Sample preparation

Vanadium pentoxide ($V_2O_5$, 99.6%, Sigma Aldrich), hydrazine hydrate solution ($N_2H_4 \cdot H_2O$, 85%, Fisher Scientific), Sulfuric acid ($H_2SO_4$, 95%, Sigma Aldrich), Cetyl Trimethy Ammonium Bromide (CTAB, 98%, Sigma Aldrich), Sodium Hydroxide (NaOH, Fisher Scientific), Ethanol ($C_2HOH$, Fisher Scientific), Low Density Polyethylene (LDPE), IRGANOX (BASF) were used as received without further purification.

*Synthesis of Vanadium dioxide ($VO_2$) rods (Figure S1.a1):* The synthesis of rods consist of two-step processes as reported earlier.[1] 0.48 g of $V_2O_5$ was dispersed in 10 mL of deionized water with continuous magnetic stirring to form a yellow suspension. 0.75 mL of $H_2SO_4$ was added while heating the suspension at 80 °C. After 15 min, 270 μL of $N_2H_2 \cdot H_2O$ was slowly added to the above mixture to form a transparent blue color solution indicating the reduction of $V^{5+}$ to $V^{4+}$. With vigorous stirring, 1.8 mL of CTAB (0.1M) was added and the solution kept stirring for 1 hour. The pH of the resulting solution was then adjusted to 4.0 - 4.2 by adding NaOH (1M) dropwise. The brown precipitate was washed with water three times by using a centrifuge and re-dispersed in 19 mL water and transferred to a 45mL Teflon Autoclave (Parr acid digestion vessel). Hydrothermal reaction was carried out at 220 °C degree for 63 hr. The final black precipitate was collected by centrifugation, washed with ethanol, and dried at 80 °C for 1hr. Thermal transformation of $VO_2$ (A) to $VO_2$ (M) was carried out in a vacuum tube furnace at 550 °C, 0.2 mbar for /1hr.

*Synthesis of Vanadium dioxide ($VO_2$) stars (Figure S1.b1):* The synthesis of stars follows a similar hydrothermal procedure than rods, except with higher $N_2H_2 \cdot H_2O$ concentration (330 μL) and higher reaction temperature (235°C degree for 63 hr). This reaction condition led to the single step formation of $VO_2$ (M) stars, which were collected by centrifugation, washed with ethanol, and dried at 80 °C for 1hr.

All hydrothermally synthesized samples were washed thoroughly with ethanol to remove excess of capping agent and reducing agent.

## S1.2. Sample Characterization and optical measurements

Morphology and size of the $VO_2$ samples were obtained using a field-emission Scanning Electron Microscope (SEM), JEOL JSM-6701F instrument with an accelerating voltage of 5-10 KeV. For SEM imaging, a small portion was drop-casted on silicon wafer followed by room temperature drying. The average particle size was measured through SEM images using ImageJ software.[2] The crystallographic phase identification of $VO_2$ samples and phase transformation with temperature were determined by powder X-ray diffraction (PXRD), using a STOE SEIFERT diffractometer with angular range of 2° < 2θ < 45° and a Mo K-alpha X-ray radiation source. X-ray photoelectron spectroscopy (XPS) was carried out on a Thermo Scientific K-alpha photoelectron spectrometer with a dual beam charge compensation system using monochromatic Alkα radiation. High-resolution scans were recorded for the principal peaks of C (1s), V (2p) and O (1s) at a pass energy of 50 eV. The binding energies were calibrated with respects to O1s peak at 530.0 eV. All peak fittings were carried out using CasaXPS software. Differential scanning calorimetry (DSC) analysis of as synthesized $VO_2$ samples were performed on a DSC instrument from Mettler Toledo, where experiments were carried out between 25 to 200 °C under nitrogen atmosphere with a heating ramp of 5 °C/min and cooling ramp of 0.5 °C/min. Thermal camera images of $VO_2$ antenna composites on aluminum foil were taken in FLIR A655c High resolution LWIR camera at room temperature and high temperature. The absorption spectra of $VO_2$ antenna composites were measured by Fourier transform Infrared (FTIR) spectroscopy (SCHIMADZU IRTracer -100) with a gold mid-IR integrating sphere (Pike Technologies), and a custom-built thermoelectric temperature controller.

### S1.3. $VO_2$ polymer composite preparation

A known amount of $VO_2$ (M) rods or stars (0.002g-0.02g) was mixed with 0.8g LDPE and 0.01g of antioxidant (IRGANOX) through solid grinding. This mixture was then poured b/w aluminum plates and subjected to compress molding at 180°C/ 350 bar for 3 min. The thickness of the polymer composite film was measured through electronic digital micrometer and cross-sectional SEM which showed average thickness of 80 µm (Figure S3). The measured emission spectra of the composite films with different rod(star) volume fractions are shown in Figures S1a2 and S1a3(S1b2 and S1b3), and Figure 1e(3b) of the main text. The volume fraction was estimated from the $VO_2$/LDPE weight ratio and the densities of $VO_2$(M) (4.230 g/cm$^3$)[3] and LDPE (0.925 g/cm$^3$).

### S1.4. $VO_2$ composite Spray coating

The $VO_2$ powder was grounded using mortar and pestle and mixed with acetone to prepare a 2 wt.% suspension. The mixture was left in an ultra-sonication bath for 4 hr to achieve a homogenous suspension. Later, the suspension was sprayed on aluminum foil using a spray gun (DeVilbiss DAGR Airbrush, Fluid tip: 0.35mm, Operating pressure: 1.3 – 3.5 bar, 10 ml suspension for 2 ×2 cm2 sample) from 15 cm at an angle of 45°. The temperature of the aluminum substrate was maintained at 50° C to ensure the rapid evaporation of any residual solvent and the formation of a smooth film. Finally, the composite samples were prepared by dip-coating the spray-coated $VO_2$ film in 2 wt% polyethylene solution in Toluene (Figure S4).

### S1.5. Orientation-averaged scattering simulations

Orientation-and-polarization-average light scattering simulations were performed using the SCUFF-EM[4] application AVESCATTER.[5] SCUFF-EM, is an open-source software for electromagnetic simulations, based on the Boundary Elements Method (BEM). The meshing of the objects is based on triangular panels and was carried by GMSH.[6] The volume of each structure was numerically computed by GMSH.

### S1.6. First principle radiative energy transfer simulations

Radiative transfer simulations were performed by mc-photon, our open-source software for Monte Carlo simulations of unpolarized light.[7] The code has been validated previously.[8] The Monte Carlo algorithm consist on simulating the trajectories of many individual photons as they interact with particles and interfaces, until they are either, absorbed by particles or exit the simulation domain. The initial condition of each photon is given by the position and direction of the light source. At each simulation step, the optical path ($\Lambda_{photon}$) and fate of a photon is estimated by selecting the shortest path between the particle's scattering ($\Lambda_{sca}$) and absorption ($\Lambda_{abs}$), the absorption of the host ($\Lambda_{host}$), or diffraction ($\Lambda_{Fresnel}$), where:

$$\Lambda_{sca} = -\frac{V_p}{f_v \langle C_{sca} \rangle} \ln \xi,$$

$$\Lambda_{abs} = -\frac{V_p}{f_v \langle C_{abs} \rangle} \ln \xi,$$

$$\Lambda_{host} = -\frac{\lambda}{4\pi \kappa_{host}} \ln \xi,$$

and $\Lambda_{Fresnel}$ is given by the shortest distance between the photon and an interface. In the equations above, $\xi$ is a random number between 0 and 1, and $\kappa_{host}$ is the imaginary part of the refractive index of the host.

If diffraction occurs ($\Lambda_{photon} = \Lambda_{Fresnel}$), the photon is either reflected or transmitted by random selection, with the probabilities of each event proportional to the respective energy flux defined by Fresnel laws. If the photon is absorbed by a particle ($\Lambda_{photon} = \Lambda_{abs}$) or the host material ($\Lambda_{photon} = \Lambda_{host}$), the event is terminated, and the simulation continues with a new photon at the initial conditions. For a scattered photon ($\Lambda_{photon} = \Lambda_{sca}$) the new direction, $\theta$, is determined by:[9]

$$\cos \theta = \begin{cases} \frac{1}{2g}\left[1 + g^2 - \left(\frac{1-g^2}{1-g+2g\xi}\right)^2\right] & \text{if } g \neq 0 \\ 2\xi - 1, & \text{if } g = 0 \end{cases}$$

where $g = \langle \mu_{sca} \rangle$, and $\mu_{sca}$ is the asymmetry parameter.

In all simulations, we considered a slab with large surface area to represent a 2D problem. As a criterion, we selected the smallest surface area by which no photon escapes through the edges. Two large monitors above and below the slab measure the total reflectance and transmittance, respectively. In all the simulation, we considered 1,000,000 photons per wavelength.

For samples featuring polydisperse particle-size distributions, $\langle C_{abs} \rangle$, $\langle C_{sca} \rangle$ and $\langle \mu_{sca} \rangle$ represent the ensemble averaged values weighted by a two-dimensional normal distribution (see details in section S2.3).

## S2. Supplementary Text

### S2.1. Characterization of VO₂ antennas

The phase purity of the hydrothermally synthesized and thermally annealed product were analyzed through PXRD. The observed diffraction peaks for the as-synthesized VO$_2$ rods (

Figure **S5**a) can be readily assigned to the tetragonal crystalline phase (space group: p4/ncc) of VO$_2$ (A), which was in very good agreement with literature values (JCPDS card no. 42–0876). Strong intensity peaks of (110), (102), and (220) at 14.80, 25.50, and 29.90° indicate good crystallinity and high purity of the as-synthesized VO$_2$ (A) phase. The phase transformation of this sample from VO$_2$ (A) to VO$_2$ (M) was carried by annealing the sample under vacuum (0.1 mbar) at 550 °C and the transformation was confirmed through PXRD peaks (

Figure **S5**b) which are in very good agreement with the literature values (JCPDS card no.43-1051). Single step hydrothermal synthesis of VO$_2$ stars sample XRD peaks were also in good agreement with standard monoclinic VO$_2$ (M) phase.

XPS measurements were carried out on the VO$_2$ samples to determine the surface composition and oxidation states. Due to their proximity, both the V 2p and the O 1s peaks were collected and fitted together. All data was charge corrected to the lattice oxygen 1s peak at 530.0 eV.[10] V 2p$_{3/2}$ peaks were centered at 516.3 and 516.2 for the micro rods and micro stars, respectively matching the V$^{4+}$ oxidation state (Figure S6).[10] Secondary peaks associated with V$_2$O$_5$ arising from surface oxidation was observed in all the samples. The V 2p$_{3/2}$ peaks for V$^{5+}$ were situated at 518.1 eV for the micro rods and at 517.8 eV for the micro stars (Table S1).[10] The presence of the V$^{5+}$ spices are attributed to surface oxidation of the samples and has been observed previously in VO$_2$ powders and thin films.[11]

**Table S1** The XPS results for the VO2 samples grown via hydrothermal methods. Peak positions and V(IV) and V(V) concentrations are given. All peaks were charge corrected to lattice oxygen at 530.0 eV.

|  | Binding energy / eV | | Concentration / at.% | |
| --- | --- | --- | --- | --- |
| **Sample** | **V(IV) 2p$_{3/2}$** | **V(V) 2p$_{3/2}$** | **V(IV)** | **V(V)** |
| **Micro rods** | 516.3 | 518.1 | 80.5 | 19.5 |
| **Micro stars** | 516.2 | 517.8 | 70.6 | 29.4 |

The thermal characterization of VO$_2$ (M) rods through DSC shows endothermic phase transition ($T_c$) at 70.2 °C as represented in Figure S7. Similar trend was observed for VO$_2$ stars.

### S2.2. Emissivity contrast formula

Using Kirchhoff's law,[12] the emissivity of a composite is given by:

$$\epsilon = 1 - \rho^* - \tau^*$$

where $\rho^*$ and $\tau^*$ are, respectively, the reflectance and transmittance of the composite.

At small concentration of antennas, the effect of scattering is negligible, and the reflectance of the hot ($\rho_h^*$) and cold ($\rho_c^*$) composite can be approximated by the reflectance of the empty host $\rho$.[13] On the other hand, through Beer-Lamberts law:[13]

$$\tau^* = (1-\rho)e^{-\left(f_v \frac{C_{abs}}{V_p} + 2\alpha_0\right)t_{film}},$$

where $\alpha_0$ is the absorption coefficient of the host material, $t_{film}$ is the thickness of the composite and $f_v$ and $V_p$ are, respectively, the volume fraction and the volume of the antennas.

Under these assumptions, we can approximate the emissivity of a composite as:

$$\epsilon = (1-\rho)\left[1 - e^{-\left(f_v \frac{C_{abs}}{V_p} + 2\alpha_0\right)t_{film}}\right] \tag{S1}$$

Using Eq. (S1) and $\Delta\epsilon = \epsilon_h - \epsilon_c$, we obtain:

$$\Delta\epsilon = \tau_c e^{-2\alpha_0 t_{film}}\left[1 - \left(\frac{\tau_c}{1-\rho}\right)^{C_{abs,h}/C_{abs,c}-1}\right], \tag{S2}$$

where $\tau_c = (1-\rho)e^{-2f_v \frac{C_{abs,c}}{V_p} t_{film}}$.

The accuracy of Eq. (S1) against full radiative transfer simulations using Monte Carlo method can be visualized in Figure S13. As shown in the figure, the discrepancy becomes significant for particles with large $\langle C_{sca}\rangle/V_p$.

### S2.3. Orientation-averaged scattering and absorption of polydisperse VO₂ rods.

First, we computed the orientation-averaged light scattering parameters of a VO₂ rod of width $W$ and length $L$, $\langle C_{abs}\rangle_{W,L}$, $\langle C_{sca}\rangle_{W,L}$ and $\langle \mu_{sca}\rangle_{W,L}$, in the range $W \in [0.05, 1.4]$ μm and $L \in [1, 50]$ μm. Each rod was simulated at hot and cold phases (Figure S11). Afterwards, the effect of size polydispersity was considered by computing the ensembled averaged $\langle C_{abs}\rangle$, $\langle C_{sca}\rangle$ and $\langle \mu_{sca}\rangle$:

$$\langle C_{abs}\rangle = \sum_{W,L} F_{W,L}\langle C_{abs}\rangle_{W,L},$$

$$\langle C_{sca}\rangle = \sum_{W,L} F_{W,L}\langle C_{sca}\rangle_{W,L},$$

$$\langle \mu_{sca}\rangle = \frac{1}{\langle C_{sca}\rangle}\left[\sum_{W,L} F_{W,L}\langle C_{sca}\rangle_{W,L}\langle \mu_{sca}\rangle_{W,L}\right],$$

where $F_{W,L}$ is a 2D Gaussian distribution:[14]

$$F_{W,L} = \frac{1}{2\pi\sigma_W\sigma_L\sqrt{1-r_{W,L}}}\exp\left\{-\frac{1}{2(1-\rho_{W,L}^2)}\left[\left(\frac{W-\bar{W}}{\sigma_W}\right)^2 - 2r_{W,L}\left(\frac{W-\bar{W}}{\sigma_W}\right)\left(\frac{L-\bar{L}}{\sigma_L}\right) + \left(\frac{L-\bar{L}}{\sigma_L}\right)^2\right]\right\},$$

where $\bar{x}$ and $\sigma_x$ represent, respectively, the mean and standard deviation of a variable $x$, and $r_{W,L}$ is the correlation coefficient between $W$ and $L$. Similarly, the ensemble-averaged volume of the sample $V_p$ is given by:

$$V_p = \sum_{W,L} F_{W,L} V_{W,L}$$

where $V_{W,L}$ correspond to the volume of an antenna of width $W$ and length $L$.

For the results of Figure 1e (main text), $\bar{W}, \sigma_W, \bar{L}, \sigma_L$ and $r_{W,L}$ were obtained from SEM images of the sample (Figure 1c). The gaussian distribution is plotted in Figure S8a, the ensemble averaged $\langle C_{abs} \rangle$, $\langle C_{sca} \rangle$ and $\langle \mu_{sca} \rangle$ in Figure S8b, and radiative energy transfer simulations in Figure S8c. In the results of Figure 2h (main text), $r_{W,L} = 0$.

## S2.4. T-Matrix mode decomposition

In the T-matrix method, the scattered field is expressed by a linear expansion into spherical vector wave functions $\vec{M}_{lm}$ and $\vec{N}_{lm}$:[15]

$$\vec{E}_{sca} = \sum_{\substack{l=1 \\ l'=1}}^{\infty} \sum_{\substack{m=-l \\ m'=-l'}}^{l,l'} \begin{bmatrix} T^{MM}_{lm,l'm'} & T^{MN}_{lm,l'm'} \\ T^{NM}_{lm,l'm'} & T^{NN}_{lm,l'm'} \end{bmatrix} \begin{pmatrix} \vec{M}_{l'm'} \\ \vec{N}_{l'm'} \end{pmatrix}$$

where $T^{MM}_{lm,l'm'}, T^{MN}_{lm,l'm'}, T^{NM}_{lm,l'm'}$ and $T^{NN}_{lm,l'm'}$ are the T-matrix modes. In a more compact form:

$$\vec{E}_{sca} = \sum_{ij}^{\infty} T_{ij} \vec{\beta}_j$$

where the indexes $i$ or $j$ are defined by the spherical harmonics indexes as:

$$i = 2[l(l+1) + m - 1] + P$$

and $P = 0, 1$ for $M$ and $N$, respectively. The following table summarizes the corresponding equivalence:

| $i$ (or $j$) | 0 | 1 | 2 | 3 | 4 | 5 | 6 | ... | 15 | 16 | ... |
|---|---|---|---|---|---|---|---|---|---|---|---|
| $l$ | 1 | 1 | 1 | 1 | 1 | 1 | 2 |  | 2 | 3 |  |
| $m$ | -1 | -1 | 0 | 0 | 1 | 1 | -2 | ... | +2 | -3 | ... |
| $P$ | 0 | 1 | 0 | 1 | 0 | 1 | 0 |  | 1 | 0 |  |

The T-matrix elements represent the contribution of electric and magnetic multipoles to the overall electromagnetic response of the structure. For example, the indexes 1, 3 and 5 (0, 2 and 4) represent the of the electric(magnetic) dipole modes in the $x, z$ and $y$ direction, respectively.[16] The T-matrix is diagonal for structures with spherical symmetry,[17] while other off-diagonal $T_{ij}$ terms appear for non-spherical structures.[18]

We performed T-matrix decomposition using the open-source application SCUFF-TMATRIX.[4] The number of modes was chosen using the formula:[19]

$$\langle C_{abs} \rangle = -\frac{2\pi}{k_0^2} \sum_{ij} \left[ \text{Re}(T_{ii})\delta_{ij} + |T_{ij}|^2 \right]. \tag{S2}$$

The equation was used to determine the minimum number of T-matrix modes, where $\langle C_{abs} \rangle$ was obtained from orientation-averaged scattering simulations.[5] The number of T-matrix modes was determined by analyzing Eq. (S2) in the wavelength range $\lambda \in [8, 13]$ µm (Figure S9b)

The results from mode decomposition reveal the mechanism behind the emissivity enhancement of the hot VO$_2$ antenna (Figure S10). For a sphere (ellipsoid with three equal axis) the response is featured by the contribution of electric and magnetic dipole modes in the $x$, $y$ and $z$. Shrinking one of the three ellipsoid axes leads to a disk-shaped particle, whose emissivity is featured by three strong electric dipole modes, one of which dominates the response. The rod shape that results when shrinking two of the three ellipsoid axes reaches the strongest emissivity, with only one strong electric dipole.

### S2.5. VO₂ core shell particles

The emissivity ratio at $\lambda = 10$ µm of core@shell sphere based on a dielectric core of refractive index $N = 1.5$ and a VO₂ shell (**Error! Reference source not found.**), was studied as a function of the core diameter ($D_c$) and shell thickness ($t_s$). For a given $D_c$ and $t_s$, a simulation was performed using the scattering theory for multilayered spheres.[20,21]

### S2.6. Analysis of the morphology of the antenna

The results shown in Figure 3e-g (main text), are based on a series of steps that include orientation-averaged scattering,[8] Monte Carlo[7] and Beer Lambert (Eq. S1) simulations. In short, for a given structure:

1. We simulated $\langle C_{\text{abs}} \rangle$, $\langle C_{\text{sca}} \rangle$ and $\langle \mu_{\text{sca}} \rangle$ (left column Figure S13).
2. These results were used as input in Monte Carlo to predict the spectral emissivity of a composite as a function of $f_v$, such as the results shown in the center column of Figure S13.
3. The emissivity at $\lambda = 10$ µm was plotted as a function of $f_v$ (right column, Figure S13).
4. The value of $f_v$ with the largest $\Delta\epsilon$ was selected and plotted in Figure 3g (main text).

# S3. Supplementary Figures

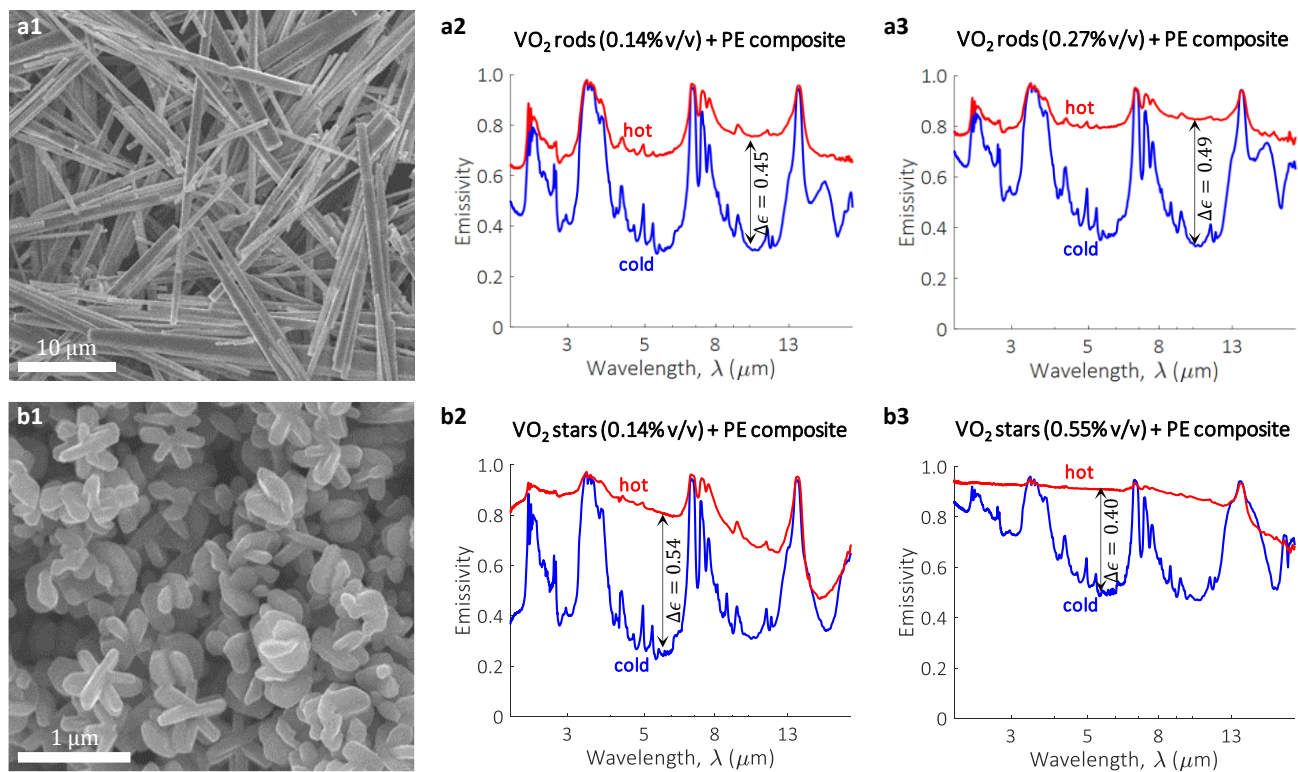

**Figure S1. a1.** and **b1.** SEM images of as-synthesized $VO_2$ rods and stars used in the results of the main text, respectively. **a2. and a3.** Measured emission spectra of $VO_2$ rods + PE composites based on the sample shown in a1, and volume fractions of 0.14% and 0.27% v/v, respectively. **b2 and b3.** Measured emission spectra of $VO_2$ stars + PE composites based on the sample shown in b1, and volume fractions of 0.14% and 0.55% v/v, respectively.

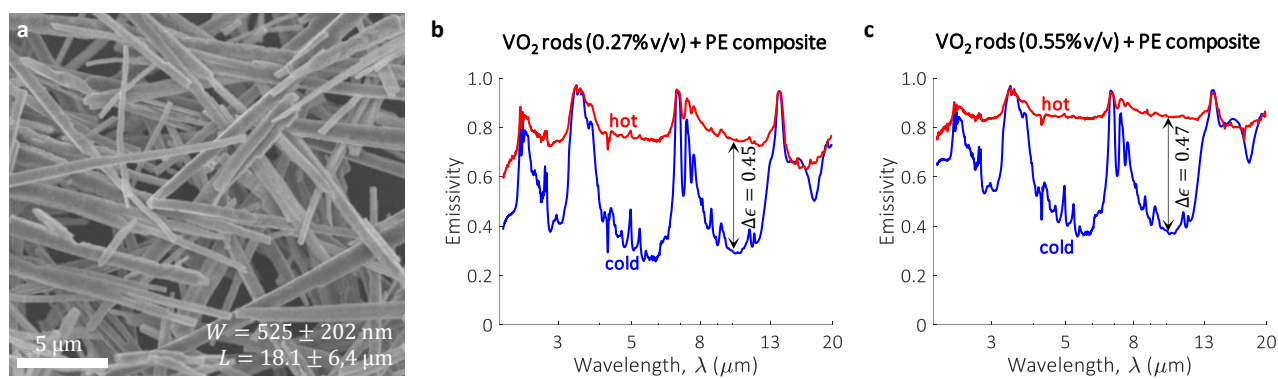

**Figure S2. a.** SEM images of another batch of $VO_2$ rods, synthesized according to the guidelines outlined in Materials and Methods (section S1.1). **b. and c.** Measured emission spectra of $VO_2$ rods + PE composites based on the sample shown in a, and volume fractions of 0.27% and 0.55% v/v, respectively.

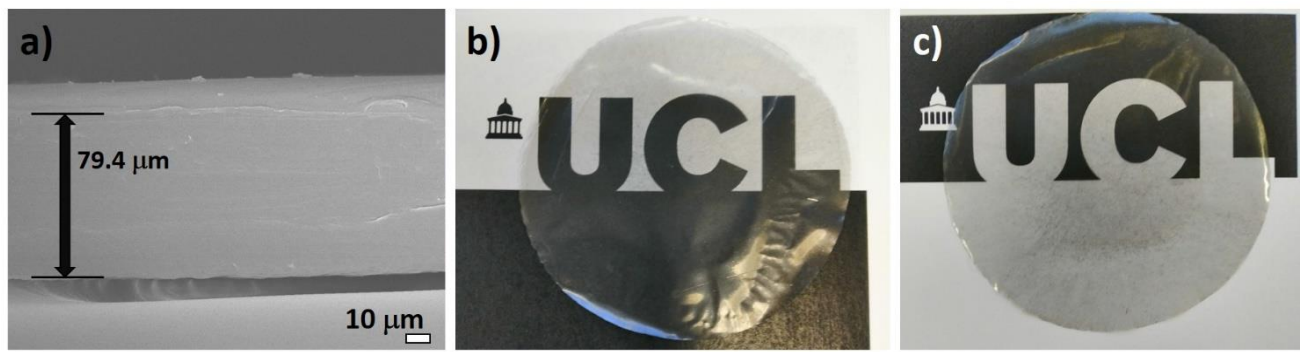

**Figure S3 a.** Cross sectional SEM image of VO$_2$ polymer composite showing the thickness of 79.4 µm. **b. & c.** Compress molded sample of polyethylene and VO$_2$ rod + polyethylene composite films.

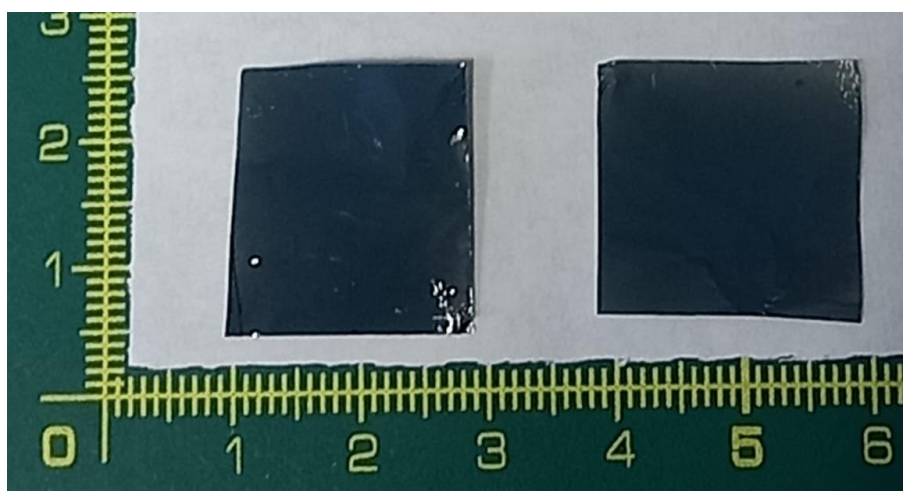

**Figure S4** Spray coated VO$_2$ rods on aluminum foil.

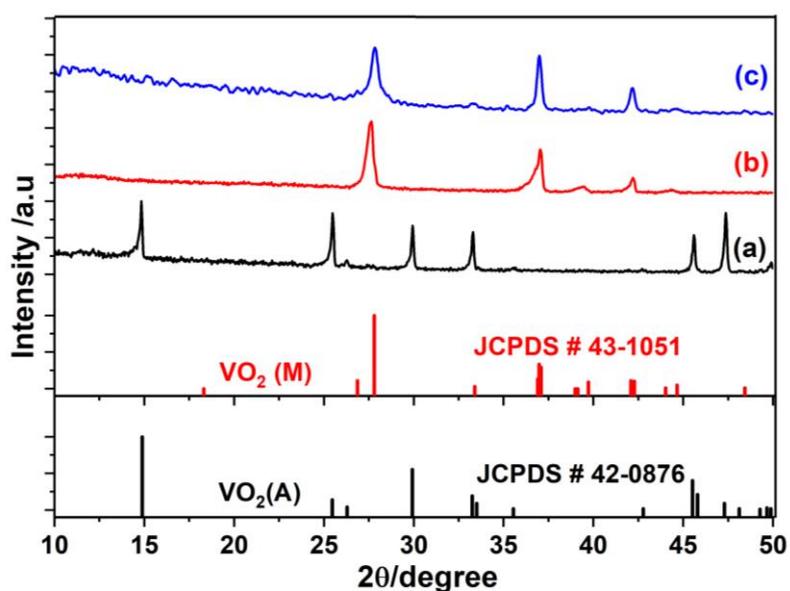

**Figure S5** Powder X-ray diffraction (PXRD) pattern; JCPDS plots of VO$_2$ (A) and VO$_2$ (M), followed by (a) as-synthesized and **(b)** annealed (550 °C) VO$_2$ rods, and **(c)** as-synthesized VO$_2$ stars.

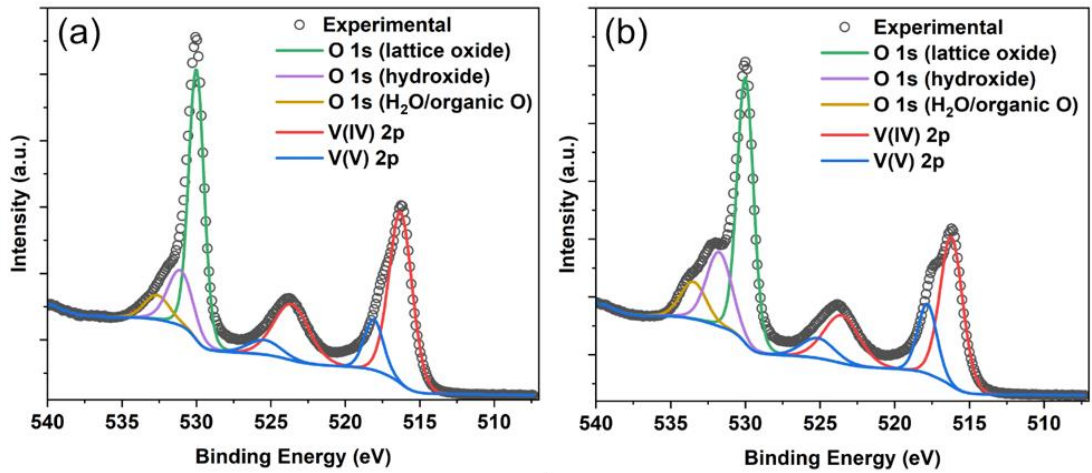

**Figure S6** High resolution V2p3/2 XPS spectra of a) VO$_2$ (M) rods and b) stars XPS data showing the O 1s and V 2p experimental (white circles) and peak fitting (colored lines). Only vanadium in the 4+ and 5+ oxidation states was detected in the sample.

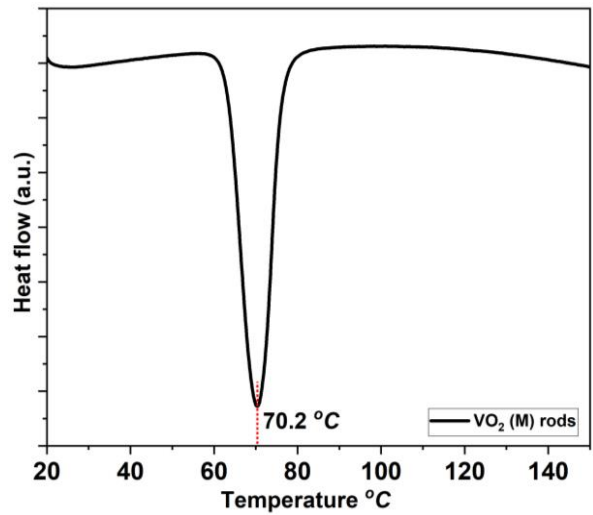

**Figure S7** Differential Scanning calorimetry (DSC) curve of VO$_2$ (M) rods showing thermochromic phase transition temperature ($T_c$) at 70.2 °C.

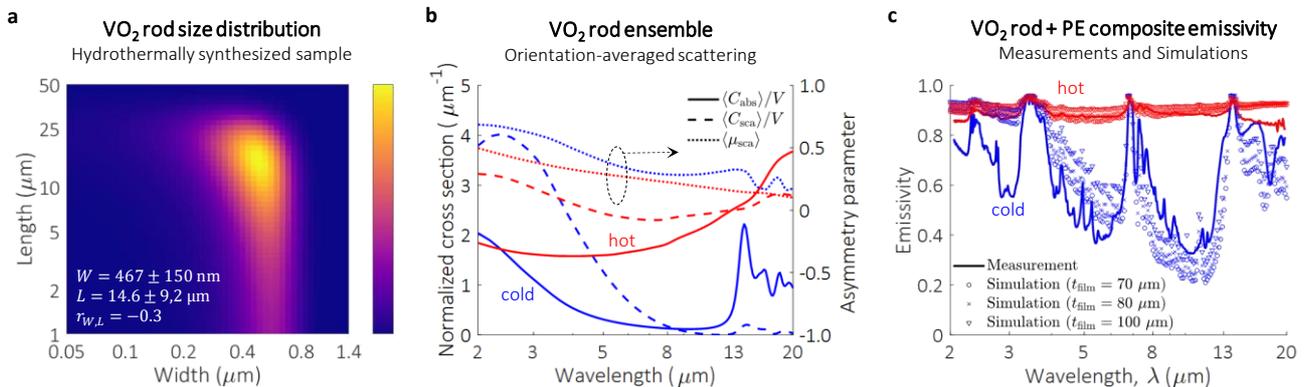

**Figure S8 a.** Size distribution of the sample of VO$_2$ rod antennas shown in Figure 1c (main text). **b.** Simulated ensemble-averaged $\langle C_{\text{abs}}\rangle/V_p$, $\langle C_{\text{sca}}\rangle/V_p$ (left vertical axis) and $\langle \mu_{\text{sca}}\rangle$ (right vertical axis) of the VO$_2$ rod antenna ensemble based the size distribution shown in (a). **c.** The results from (b) were used to simulate the emissivity of the composite and compare with the measurements shown in Figure 1e (main text). The simulation considered three film thicknesses ($t_{\text{film}}$) within the tolerance measured in the sample (80 ± 15 μm).

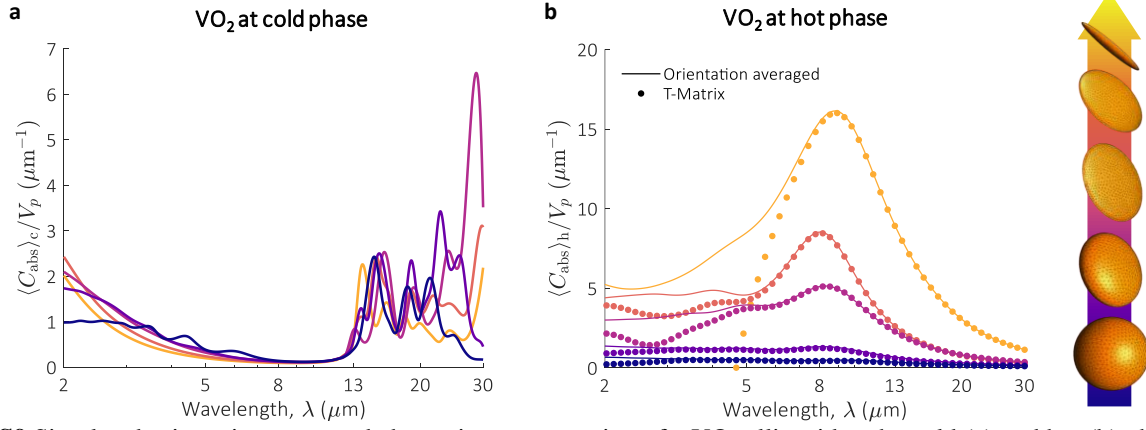

**Figure S9** Simulated orientation-averaged absorption cross section of a VO$_2$ ellipsoid at the cold (a) and hot (b) phase, as a function of SA:V. The dimensions are shown in Figure 2c (main text). In figure b, the filled circles represent $\langle C_{abs} \rangle$ computed by T-matrix mode decomposition, as a method of validation to determine the number of T-matrix modes (see Section S2.4).

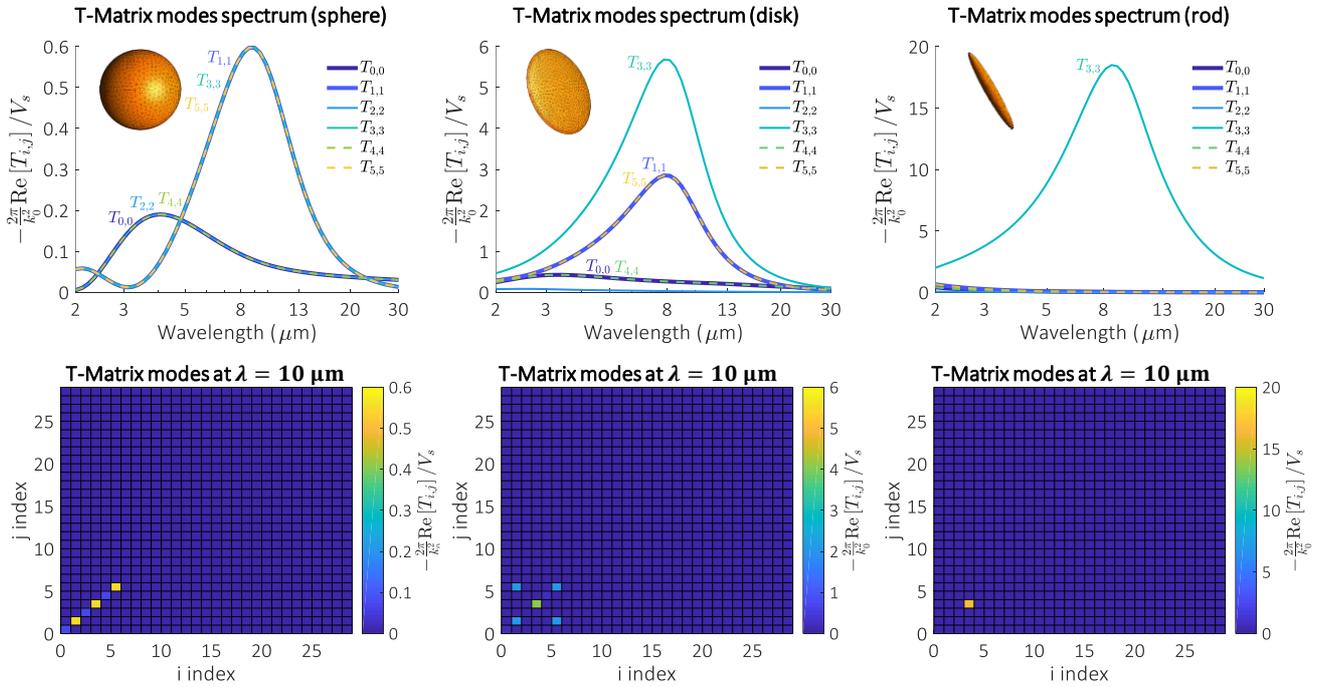

**Figure S10** T-matrix mode decomposition of a VO$_2$ particle as a function of its SA:V ratio. Results for a sphere, disk and rod shapes are shown at the left, center, and right columns, respectively. The top row shows the spectral response of the modes at the diagonal of the T-matrix, where $T_{1,1}$, $T_{3,3}$ and $T_{5,5}$ ($T_{0,0}$, $T_{2,2}$ and $T_{4,4}$) represent the electric (magnetic) dipole modes at x, z and y direction, respectively. The bottom row shows the full T-matrix at $\lambda = 10$ μm. In all the figures, the T-matrix modes are expressed as $-\frac{2\pi}{k_0^2}\text{Re}(T_{ij})/V_s$ for direct comparison, in accordance with equation (S2).

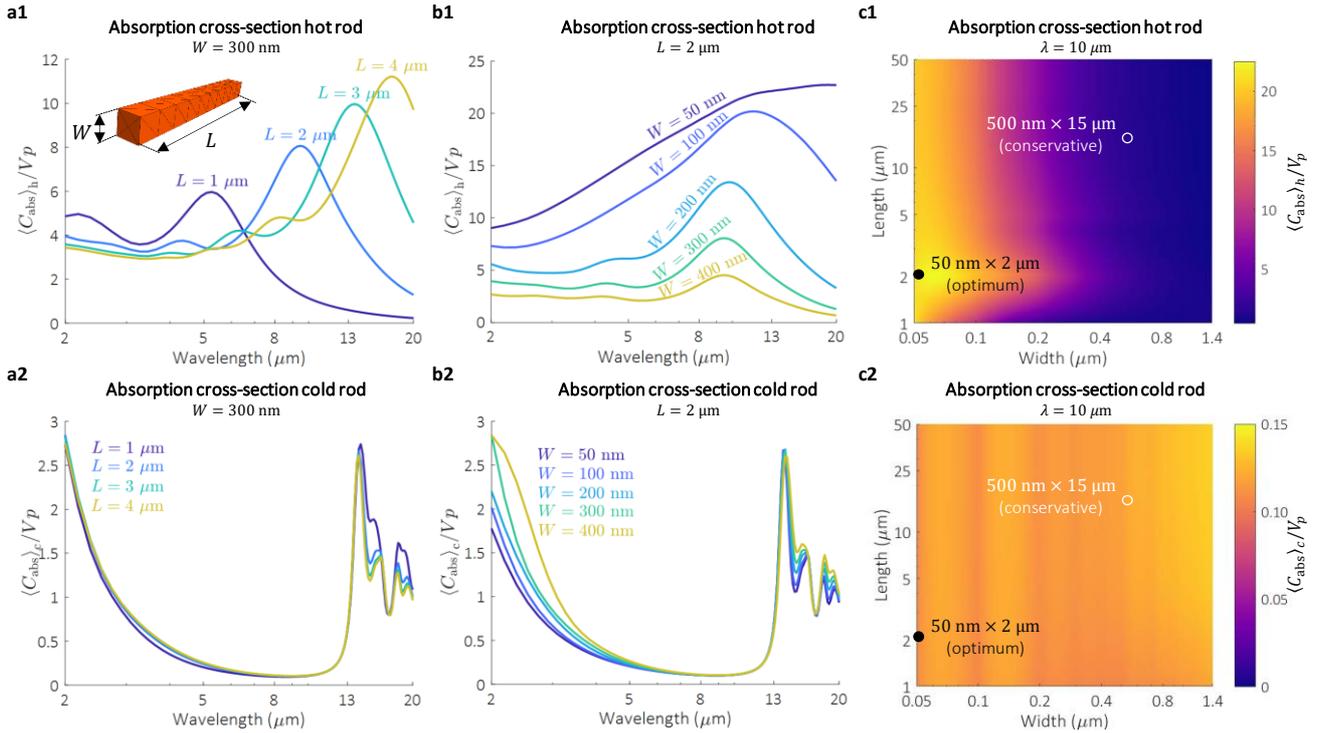

**Figure S11 a1(a2). and b1(b2).** Orientation-averaged absorption cross-section of hot(cold) phase VO$_2$ rods at as a function of $L$ and $W$, respectively. **a3(b3)** Orientation-averaged absorption cross-section of hot(cold) phase VO$_2$ rod at λ=10 μm as a function $W$ and $L$. The results shown here were used to used to elaborate Figures 2d-e of the main text.

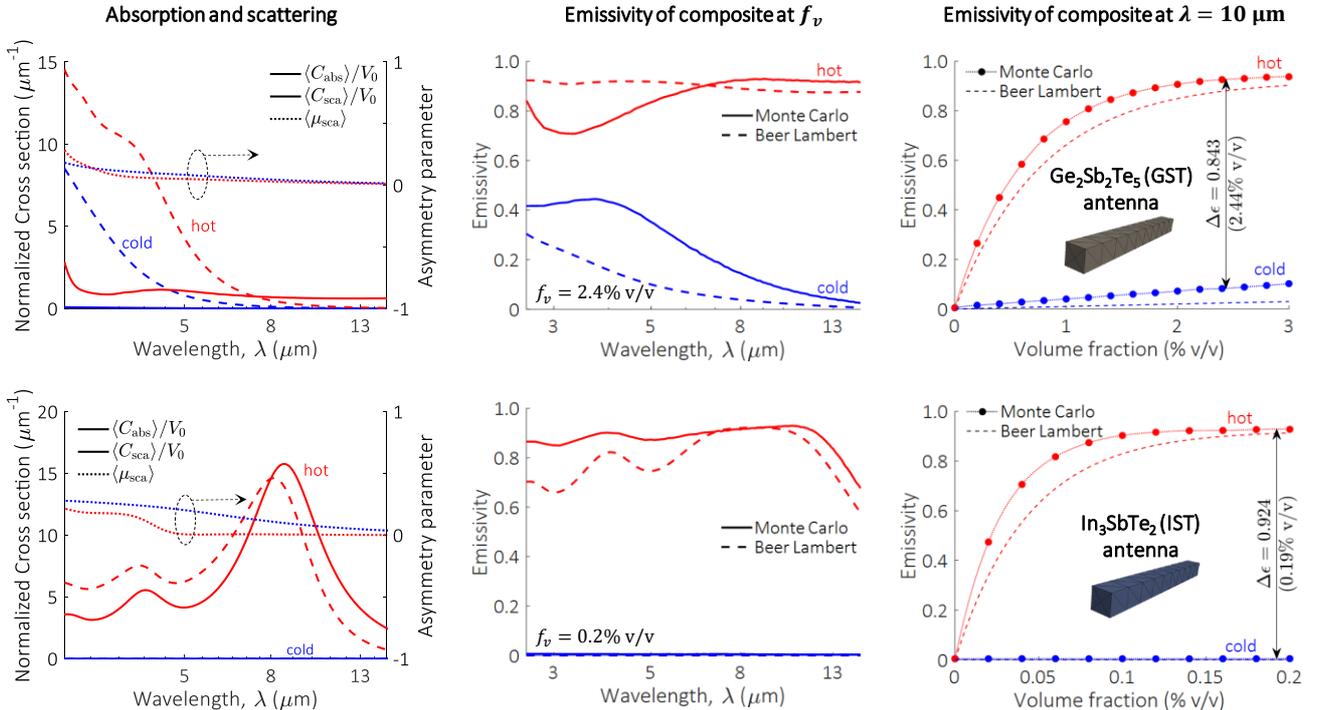

**Figure S12** Analysis of GST (top row) and IST (bottom row) rod antennas and composites. Each row shows **(left)** normalized scattering and absorption cross section (left vertical axis), and asymmetry parameter (right vertical axis); **(center)** spectral emissivity of a composite; and **(right)** composite emissivity at $\lambda = 10$ μm against volume fraction. In all simulations, the refractive index of the host, $N_h = 1.5$. The refractive index of GST and IST is reported elsewhere.[23,24] In all simulations, the rod width and length is 200 nm and 2 μm, respectively. The results from Beer-Lambert are obtained on Eq. S1 and $\langle C_{abs}\rangle/V_p$. For the two figures at the right column, each circle corresponds to a Monte Carlo simulation, and the dotted lines are drawn as a guide to the eye.

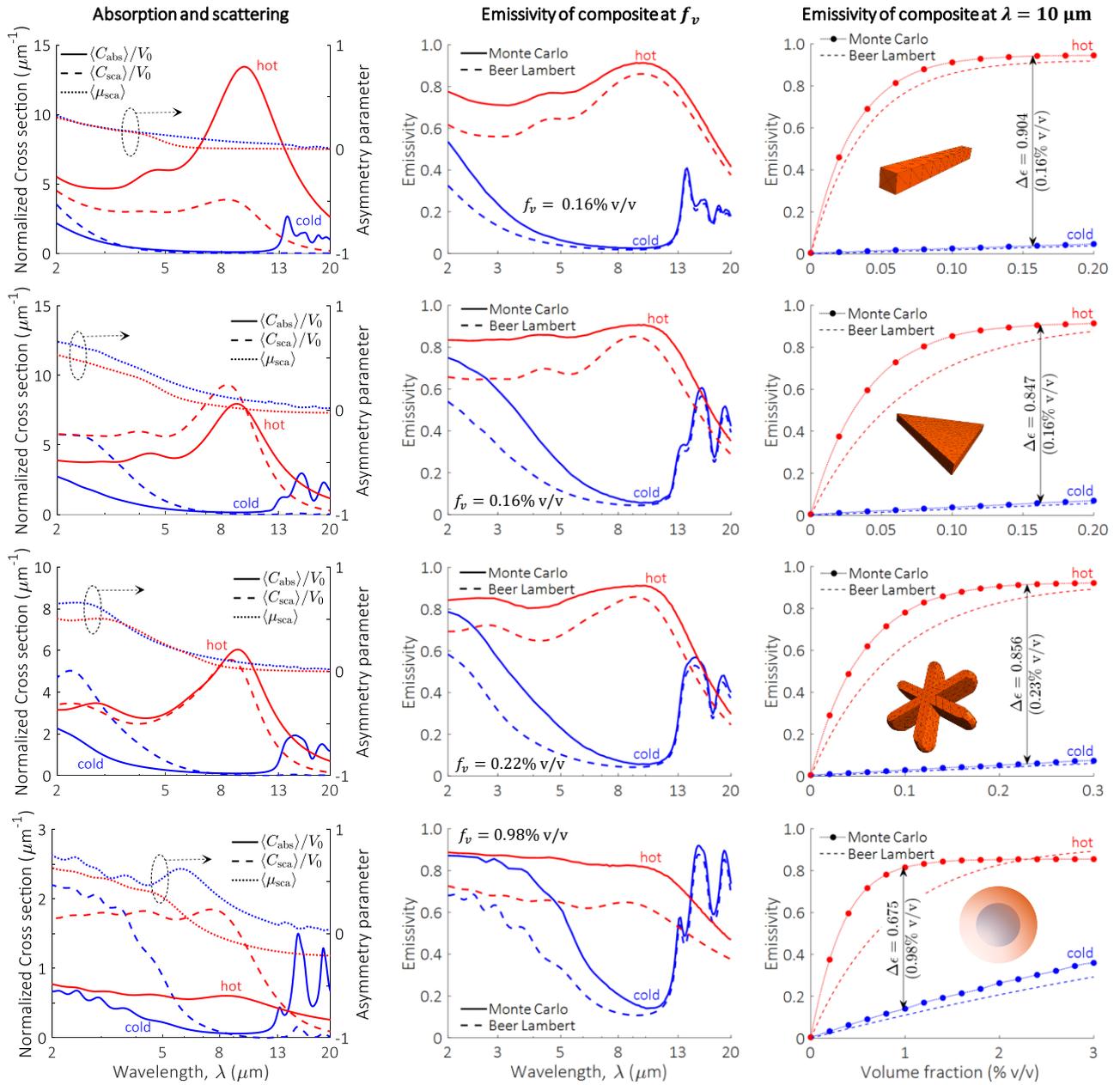

**Figure S13** (From first row) Simulation of rod, flake and star-shaped VO$_2$ antennas and core-shell particle with a VO$_2$ shell (at each row), showing **(left)** normalized scattering and absorption cross section (left vertical axis), and asymmetry parameter (right vertical axis)**; (center)** spectral emissivity of a composite; and **(right)** composite emissivity at $\lambda$ = 10 μm against volume fraction. In all simulations, the refractive index of the host, $N_h = 1.5$, and the refractive index of VO$_2$ was obtained from ellipsometry of VO$_2$ thin films over Al$_2$O$_3$.[22] The dimensions of each structure are shown in Fig. 3d (main text). The results from Beer-Lambert are obtained from Eq. S1 and $\langle C_{abs}\rangle/V_p$. For the figures at the right column, each circle corresponds to a Monte Carlo simulation, and the dotted lines are drawn as a guide to the eye**.**